\documentclass[reprint,showpacs,superscriptaddress,amsmath,amssymb,aps,prb,floatfix,]{revtex4-1}

\usepackage{graphicx}
\usepackage{hyperref}

\newcommand{\deriv}[2]{ \frac{\mathrm{d}#1}{\mathrm{d}#2} }
\newcommand{\pderiv}[2]{ \frac{\partial #1}{\partial #2} }
\newcommand{\etal}{{\it et al.\ }}

\begin{document}

\title{Cycloidal vs.\ skyrmionic states in mesoscopic chiral magnets}
\author{Jeroen Mulkers}
\email[Email: ]{jeroen.mulkers@uantwerpen.be}
\affiliation{Department of Physics, Antwerp University, Antwerp, Belgium}
\affiliation{DyNaMat Lab, Department of Solid State Sciences, Ghent University, Ghent, Belgium}
\author{Milorad V. Milo\v{s}evi\'c}
\affiliation{Department of Physics, Antwerp University, Antwerp, Belgium}
\author{Bartel Van Waeyenberge}
\affiliation{DyNaMat Lab, Department of Solid State Sciences, Ghent University, Ghent, Belgium}
\date{\today}

\begin{abstract}
    When subjected to the interfacially induced Dzyaloshinskii-Moriya interaction, the ground state in thin ferromagnetic films with high perpendicular anisotropy is cycloidal. The period of this cycloidal state depends on the strength of the Dzyaloshinskii-Moriya interaction. In this work we have studied the effect of confinement on the magnetic ground state and excited states, and we determined the phase diagram of thin strips and thin square platelets by means of micromagnetic calculations. We show that multiple cycloidal states with different periods can be stable in laterally confined films, where the period of the cycloids does not depend solely on the Dzyaloshinskii-Moriya interaction strength but also on the dimensions of the film. The more complex states comprising skyrmions are also found to be stable, though with higher energy.
\end{abstract}

\pacs{75.70.-i,75.70.Ak,75.70.Kw,75.78.Cd}
\keywords{Dzyaloshinskii-Moriya interaction, cycloidal state, skyrmion, micromagnetism, perpendicular magnetic anisotropy}

\maketitle

\section{\label{sec:intro}Introduction}

Magnets can be chiral due to the Dzyaloshinskii-Moriya interaction (DMI). In bulk materials, this interaction is caused by a lack of inversion symmetry in the crystal structure,\cite{Dzyaloshinsky1958,Moriya1960,Dzyaloshinsky1964} but in thin films DMI can also be induced by symmetry breaking at interfaces.\cite{Crepieux1998} Bogdanov \textit{et al.} gave a micromagnetic description of the basic chiral spin states--helices, cycloids, skyrmions--in ferromagnetic materials subject to the DMI.\cite{Bogdanov1989,Bogdanov1994,Bogdanov2001,Rossler2006} Bode \etal imaged the cycloidal state in a single atomic layer of manganese in 2007,\cite{Bode2007} whereas the existence of skyrmions and skyrmion lattices was only confirmed experimentally in 2009.\cite{Muhlbauer2009,Neubauer2009,Heinze2011} More exotic chiral magnetic structures have been observed in a Sc-doped barium hexaferrite thin film by Yu \etal\cite{Yu2012}

Skyrmionics, and the study of related chiral spin states, has gained a lot of interest since the first experimental evidence for the existence of magnetic skyrmions. In particular, the effect of the interfacially induced DMI on the magnetization of thin films with perpendicular magnetic anisotropy (PMA) became a prominent subject in micromagnetism. As one prominent effect, DMI makes N\'eel walls energetically favorable in thin PMA films. Consequently, the magnetic ground state is no longer homogeneous but cycloidal (in the absence of an external magnetic field) or a N\'eel skyrmion lattice (in a perpendicular field).\cite{Chen2013,Chen2015} Many applications based on the DMI in thin PMA films have already been proposed: skyrmion writer\cite{Romming2013}, racetrack memory\cite{Fert2013,Iwasaki2014,Tomasello2014,Purnama2015,Zhang2015}, skyrmion-based transistor\cite{Zhang2015b}, storage\cite{Kiselev2011} and logic gates\cite{Zhang2015a}. The DMI strength can be tuned experimentally by using different substrates, different film thicknesses, or by using a Ta buffer layer.\cite{Siegfried2015,Cho2015,Kim2015,Ganguly2015} To cover a broad range of possible materials, we take the DMI strength as a variable parameter.

For prospective applications to compete with existing information technologies, the device dimensions should be small.\cite{Hoffmann2014} When entering the mesoscopic regime, i.e., when the dimensions of the system become comparable to the typical length scales of the magnetic state, one has to take into account the effect of the confinement on the magnetic structures. Keesman \etal have studied the effect of confinement on skyrmionic ground states in thin PMA strips using Monte Carlo simulations.\cite{Keesman2015} Stabilization of skyrmions in nanowires, as a function of sample parameters, has been considered in Ref. \cite{Chui2015} by solving the Landau-Lifshitz-Gilbert equation. The skyrmionic state in a mesoscopic disk was studied by Rohart \etal\cite{Rohart2013} and is to date the only analytic consideration of a confined chiral state.

In this paper, we contribute to the understanding of the phase diagram of PMA strips and square platelets in the presence of DMI. First, we present an analytic derivation of the cycloidal state in an infinitly long strip of finite width, where demagnetization is approximated by an effective anisotropy, and the effect of the boundaries is carefully discussed. Subsequently, the analytic results are compared with micromagnetic simulations, where we do not use an approximation for the demagnetization energy. This enables us to check the validity of the effective anisotropy approximation for finite-size thin PMA films. Finally we determine the ground state and excited states of thin square PMA platelets of different sizes, and we present a complete equilibrium phase diagram and its governing rules.

The paper is organized as follows. Section~\ref{sec:methods} presents the micromagnetic framework and numerical algorithms. Section.~\ref{sec:strip} presents our results for cycloidal states in long mesoscopic strips, considered analytically (Sec.~\ref{sec:analytic}) and numerically (Sec.~\ref{sec:simul}). In Sec.~\ref{sec:square}, we finally present the complete diagram of cycloidal and skyrmionic phases for a square platelet of varied size and DMI. A summary is given in Sec.~\ref{sec:conclusions}.

\section{\label{sec:methods}Methods}

In this section, we recapitulate the micromagnetic description of thin PMA films.\cite{Cimrak2007} The quantity of interest is the magnetization field~$\vec{M}(x,y)=M_{\mathrm{sat}}\vec{m}(x,y)$ with magnetization modulus~$|\vec{M}|=M_{\mathrm{sat}}$ and magnetization direction~$\vec{m}(x,y)$. The dynamics of the magnetization is governed by the Landau-Lifshitz-Gilbert equation (LLG)
\begin{equation}
    \vec{m}_t = \frac{\gamma_{\mathrm{LL}}}{1+\alpha^2} \left( \vec{m}\times \vec{H}_{\mathrm{eff}}+\alpha \left[ \vec{m}\times(\vec{m}\times \vec{H}_{\mathrm{eff}}) \right]\right).
\end{equation}
with damping factor~$\alpha$ and the gyromagnetic ratio~$\gamma_{\mathrm{LL}}$. The effective magnetic field is the derivative of the magnetic energy density~$\varepsilon$: ~$\vec{H}_{\mathrm{eff}}=- \partial\varepsilon / \partial \vec{m}$.
When studying very thin films with saturation magnetization~$M_{\mathrm{sat}}$ and anisotropy constant~$K$, one can approximate the demagnetization energy by using an effective anisotropy $K_{\mathrm{eff}}=K-1/2\mu_0 M_{\mathrm{sat}}^2$.\cite{Coey} The three remaining energy terms of interest in this paper are related to exchange, DMI and magnetic anisotropy, respectively
\begin{eqnarray}
    \varepsilon_{\mathrm{ex}} &=& {\textstyle A \left[ \left(\pderiv{\vec{m}}{x}\right)^2 + \left(\pderiv{\vec{m}}{y}\right)^2  \right]}, \label{eq:exch} \\
    \varepsilon_{\mathrm{dmi}} &=&  {\textstyle D \left[ m_x\pderiv{m_z}{x} - m_z\pderiv{m_x}{x} + m_y\pderiv{m_z}{y} - m_z\pderiv{m_y}{y} \right]} \label{eq:dmi},\\
    \varepsilon_{\mathrm{anis}} &=& -K_{\mathrm{eff}} m_z^2. \label{eq:anis}
\end{eqnarray}

The total magnetic energy~$E(\vec{m})=\int \varepsilon(\vec{m})\mathrm{d}V$ of a thin film is a functional of the magnetization direction~$\vec{m}(x,y)$. Minimizing the total energy yields the magnetic ground state (global minimum) and the excited states (local minima). In our work we use both analytic and numerical techniques to minimize the energy.
\paragraph{Analytic approach} The cycloidal state in infinite PMA films and skyrmionic states in circular PMA disks have been calculated analytically with variational calculus in Ref.~\citenum{Rohart2013}. We have used the same approach to derive the confined cycloidal state in thin PMA strips.
\paragraph{Numerical methods}
Using the LLG equation, we follow the magnetization converging to a stable state at a local energy minimum.\cite{Cimrak2007} Subtracting the Larmor precession term from the LLG equation speeds up the computation of relaxed states. Different initial conditions can be used to find different stable magnetic states. In our simulations we started from random magnetic states, Voronoi like domains and smart initial guesses to probe the equilibrium phase diagram.

We use the finite-difference-based simulation package Mumax3 for the micromagnetic simulations presented in this paper.\cite{Vansteenkiste2014} In these simulations, we calculate the demagnetization field and do not use the thin film approximation with the effective anisotropy. We employ the boundary conditions
\begin{equation}
    \deriv{\vec{m}}{n} = \frac{D}{2A} (\vec{e}_z\times\vec{n})\times \vec{m}
\end{equation}
at an edge with normal~$\vec{n}$.\cite{Rohart2013,Vansteenkiste2014} The origin of these boundary conditions will become clear in section~\ref{sec:analytic}.

We use material parameters corresponding to Pt/Co films, as used in Ref.~\citenum{Sampaio2013}: $M_{\mathrm{sat}}$=580~kA/m, $A$=15~pJ/m, $K$=0.8~MJ/m$^3$ and $K_{\mathrm{eff}}=0.59$~MJ/m$^3$. The thickness of the film~$t$ is 0.4~nm (a single layer). The used cell size in all our simulations is 1~nm\,$\times$\,1~nm\,$\times$\,0.4~nm. This cell size guarantees a maximal angular variation of the magnetization in neighboring cells below 20$^\circ$ while preserving a reasonable computation time.

\section{\label{sec:strip}Cycloidal state in a magnetic strip}

\subsection{\label{sec:analytic}Analytic considerations}

\begin{figure}[h]
\includegraphics{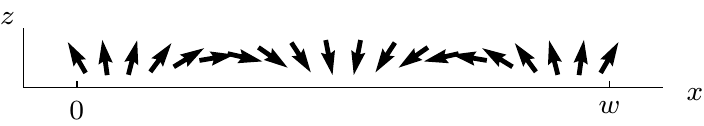}
\caption{\label{fig:drawing}Cycloidal state in a strip of width~$w$.}
\end{figure}

The first steps in the derivation of the cycloidal state in thin strips are analogous to the derivation of the cycloidal state in infinite films presented in Ref.~\citenum{Rohart2013}. The cycloidal state in a thin PMA strip (with infinite length but with finite width~$w$) can be calculated analytically after assuming that the magnetization of the cycloidal state rotates in the $(x,z)$-plane and changes only along the width of the strip~($x$-direction) and is thus constant along the length ($y$-direction) and the height ($z$-direction). This means that the magnetization $m$ is now fully described by a single angle~$\theta(x)$: $\vec{m}=(\sin(\theta),0,\cos(\theta))$. Figure~\ref{fig:drawing} sketches the magnetization in a thin PMA strip.

One can make the total energy functional, given in section~\ref{sec:methods}, more explicit by using these assumptions about the magnetization of the cycloidal state. In this paper we will work with the average energy density functional $\mathcal{E}$ which is proportional to the total energy~$E$:
\begin{equation}
    \mathcal{E}\left[\theta\right] = \frac{1}{w} \int_0^w \left[ A\left(\deriv{\theta}{x}\right)^2 - D\deriv{\theta}{x} - K_{\mathrm{eff}}\cos^2 \theta \right] \text{d} x.
\end{equation}
Here we assume free boundary conditions at $x=0$ and $x=w$. Using variational calculus we minimize (or maximize) the energy functional~$\mathcal{E}[\theta]$. This gives the Euler-Lagrange equation
\begin{equation}
    \deriv{^2\theta}{x^2} = \frac{K_{\mathrm{eff}}}{A} \sin\theta\cos\theta  \quad \text{for } 0 < x < w.
\end{equation}
The free boundary conditions become Dirichlet boundary conditions after the minimization:
\begin{equation}
    \left. \deriv{\theta}{x} \right|_{x=0} = \left. \deriv{\theta}{x} \right|_{x=w}  = \frac{D}{2A}. \label{eq:bc}
\end{equation}
Taking the indefinite integral and subsequently the square roots of both sides yields
\begin{equation}
    \deriv{\theta}{x} = \pm\sqrt{\frac{K_{\mathrm{eff}}}{A}} \sqrt{C+\sin^2\theta}.\label{eq:fdeq}
\end{equation}
Later on, we will use the integration constant~$C$ as the tuning parameter for meeting the boundary conditions. Eq.~(\ref{eq:fdeq}) tells one that the angle~$\theta(x)$ is a monotonic function. When looking at the energy functional~$\mathcal{E}$, especially at the sign of the DMI term, one can conclude that the angle~$\theta(x)$ is a monotonically increasing function for a magnetic state with a local energy minimum. This is why we will only consider the positive square root of Eq.~(\ref{eq:fdeq}). Inverting and integrating Eq.~(\ref{eq:fdeq}) from 0 to $x$ yields
\begin{equation}
    \label{eq:integral}
    \sqrt{\frac{K_{\mathrm{eff}}}{A}} x =  \int_{\theta_0}^{\theta(x)} \frac{1}{\sqrt{C+\sin^2\theta}} \text{d}\theta,
\end{equation}
with the yet unknown initial angle~$\theta_0 := \theta(0)$. This is an implicit expression for the angle~$\theta(x)$. One can calculate the quarter period of the cycloidal state by integrating the integrand in Eq.~(\ref{eq:integral}) from 0 to $\pi/2$. The period of the ground state in an infinite film can be found by altering the integration constant~$C$ until the energy density of the corresponding magnetic state~$\theta(x)$ reaches the energy minimum.

\begin{figure}[h]
\includegraphics{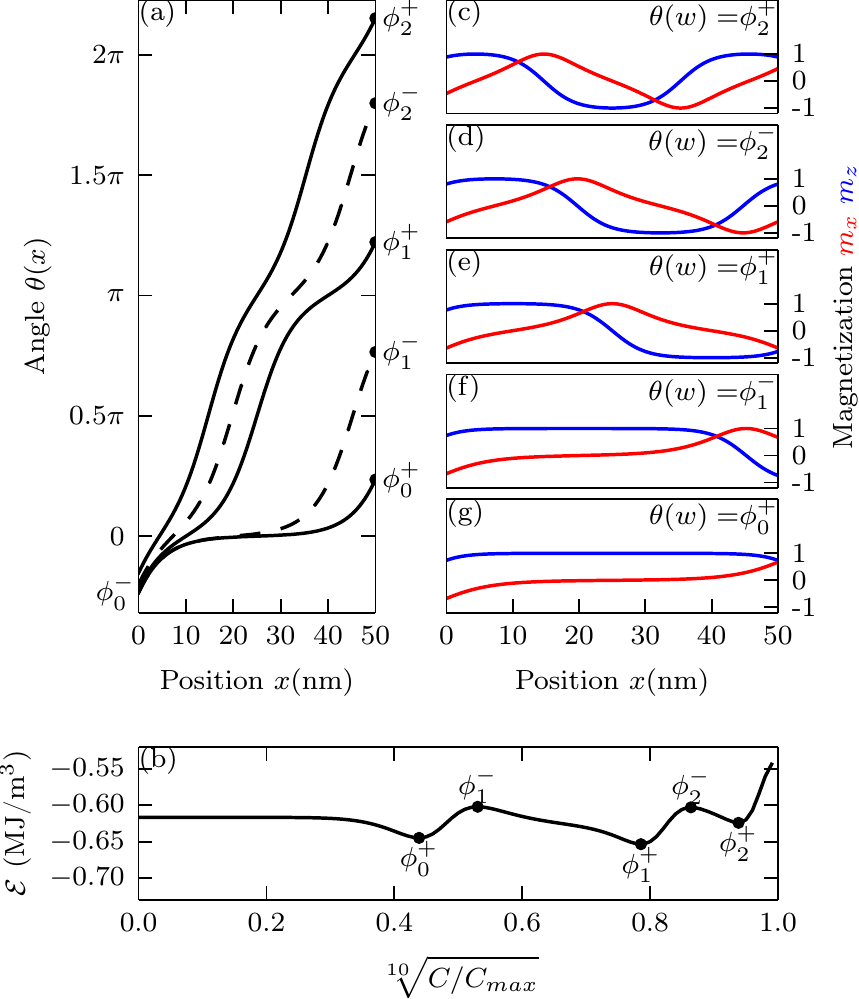}
\caption{\label{fig:sweepC}(a) $\theta(x)$ profiles of the stable cycloidal states with initial angle~$\theta_0=\phi^-_0$ in a 50~nm wide PMA strip with DMI strength~$D=4$~mJ/m$^2$. The magnetic energy density is given in function of the free parameter~$C$ in panel (b). The magnetization $m_z=\cos\theta$ and $m_x=\sin\theta$ of the stable and meta stable states are shown in (c)-(g). The stable and meta stable states are denoted by their boundary angle~$\theta(w)\in\phi^{\pm}_n$.}
\end{figure}

In contrast to the work of Rohart and Thiaville's in Ref.~\citenum{Rohart2013}, we will focus on the cycloidal state in thin strips where the boundary conditions need a special treatment. Using equations~(\ref{eq:bc}) and~(\ref{eq:fdeq}) we conclude that the angles at the boundaries have to be in the set
\begin{equation}
    \phi_n^{\pm} = \pm \arcsin\left(\sqrt{\frac{D^2}{4AK_{\mathrm{eff}}}-C}\right)+n\pi\ ,\text{with } n\in \mathbb{Z} \label{eq:possiblethetaBC}
\end{equation}
in order to meet the Dirichlet boundary conditions. From this set we choose the initial angle~$\theta_0$ (left boundary) to be $\phi^{-}_0$. The integration constant~$C$ can now be tuned in order to meet the boundary condition at $x=w$, i.e.\ until $\theta(w)\in \phi_n^{\pm}$. We do this by scanning $C$ from 0 to $C_{max} = D^2/(4AK_{\mathrm{eff}})$. For every $C$ we calculate the magnetic state $\theta(x)$ and the corresponding energy and check if $\theta(w)\in \phi_n^{\pm}$. As a representative example we discuss a full sweep of~$C$ for a strip of width~$w=50$~nm and DMI strength~$D=$4~mJ/m$^2$. The results are shown in Fig.~\ref{fig:sweepC}.

There are five different values of $C$ which yield a correct boundary angle~$\theta(w)\in\phi^{\pm}_n$ for the given example. Their corresponding energies are local energy extrema, as is expected for the Euler Lagrange equations. The three stable states have a border angle~$\theta(w)=\phi^+_n$ with $n=0,1,2$. From Fig.~\ref{fig:sweepC}(c-g) and symmetry arguments we can conclude that for stable states in general, the left boundary angle~$\theta_0\in\phi^-_n$ and the right boundary angle $\theta(w)\in\phi^+_n$. This confirms the correctness of our initial guess for the initial angle~$\theta_0=\phi_0^-$. For the given example in Fig.~\ref{fig:sweepC}, it is easy to check that the three stable states are the only stable states: choosing a different initial angle~$\theta_0\in\phi^-_n$ yields equivalent solutions due to the periodicity of~$\theta(x)$. From here on, we will label the stable states with the given integer~$n\in\mathbb{N}$. The total rotation of the magnetization of state $n$ is $n\pi$ plus a small correction in order to satisfy the boundary conditions:
\begin{equation}
    \phi^+_n - \phi^-_0 = n\pi + 2 \arcsin\left(\sqrt{\frac{D^2}{4AK_{\mathrm{eff}}}-C}\right).
\end{equation}

\begin{figure}[h]
\includegraphics{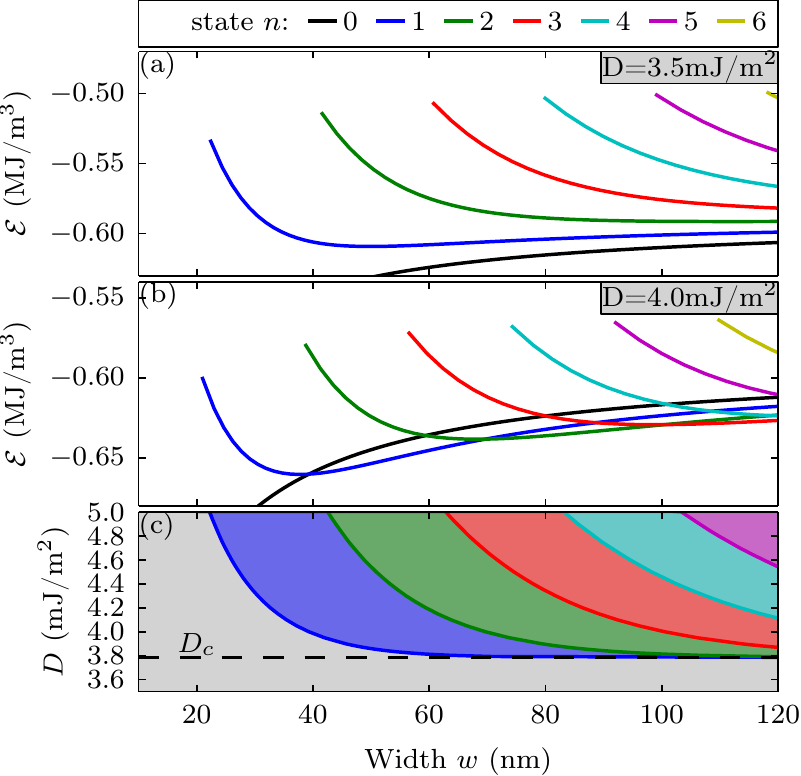}
\caption{\label{fig:cycloidgroundstates} The energy density of the stable states in a thin PMA strip is shown for DMI strength~$D=3.5$~mJ/m$^2$ (a) and $D=4.0$~mJ/m$^2$ (b). The ground state in function of the DMI interaction strength~$D$ and the width~$w$ is presented in panel (c).}
\end{figure}

We can repeat our calculations to obtain the stable states in strips of different widths~$w$ and for varying DMI strengths~$D$. The energy densities of the stable states are shown in Fig.~\ref{fig:cycloidgroundstates}(a,b) for two different DMI strengths~$D$. With these energy plots, one determines if a cycloidal state is stable in a strip of width~$w$. After finding the lowest energy state for each $D$ and $w$ we obtain the phase diagram of the ground state shown in Fig.~\ref{fig:cycloidgroundstates}(c).

If the DMI strength~$D$ is below the critical DMI strength~$D_c=4\sqrt{AK_{\mathrm{eff}}}/\pi$ then the ground state is the cycloidal state $n=0$ (the quasi uniform state). This is similar for infinite films.~\cite{Rohart2013} If $D>D_c$ then the ground state depends on the width~$w$ of the strip: the larger the width or the stronger the DMI~$D$, the larger is the $n$ of the ground state. This behavior is consistent with the results for infinite films.~\cite{Rohart2013}

\begin{figure}[h]
\includegraphics{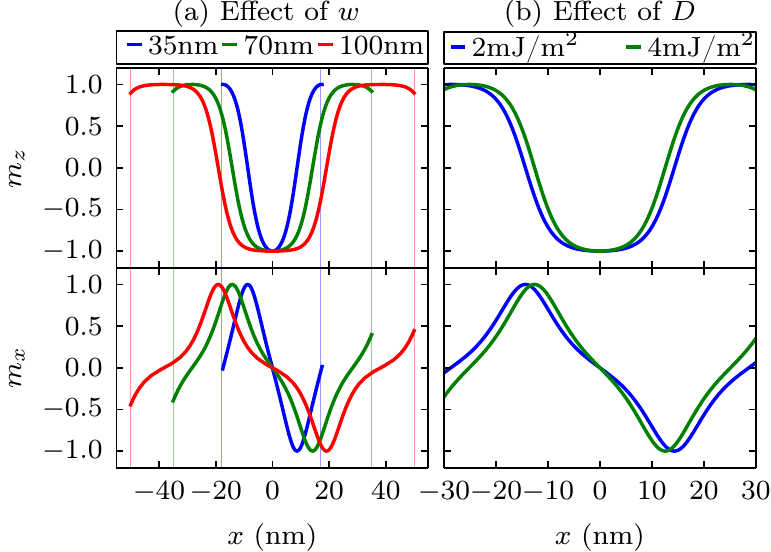}
\caption{\label{fig:varDvarL} The $z$ and $x$ component of the magnetization in the cycloidal stable state~$n=2$ in a PMA strip with width~$w$ and DMI strength~D. Panel~(a) illustrates the influence of the width~$w$ with $D=4$mJ/m$^2$ and panel~(b) the influence of the DMI strength~$D$ in a 60 nm wide strip.}
\end{figure}

The effect of the DMI and the width of the strip on a cycloidal state is shown in Fig.~\ref{fig:varDvarL}. Making the strip narrower compresses the state. It is interesting to note that $|m_z|\rightarrow1$ at the boundaries when narrowing the width~$w$ of the strip. If we confine the strip even further, then the state becomes unstable. The periodicity of the cycloidal state in infinite films strongly depends on the DMI strength~$D$. For a strip of a given width, the periodicity of a cycloidal state is practically fixed as a result of the confinement. Still, some effect of the DMI strength~$D$ is visible since it alters the boundary condition [see Eq.~(\ref{eq:bc})].

\subsection{\label{sec:simul}Micromagnetic simulations}

\begin{figure}
\includegraphics{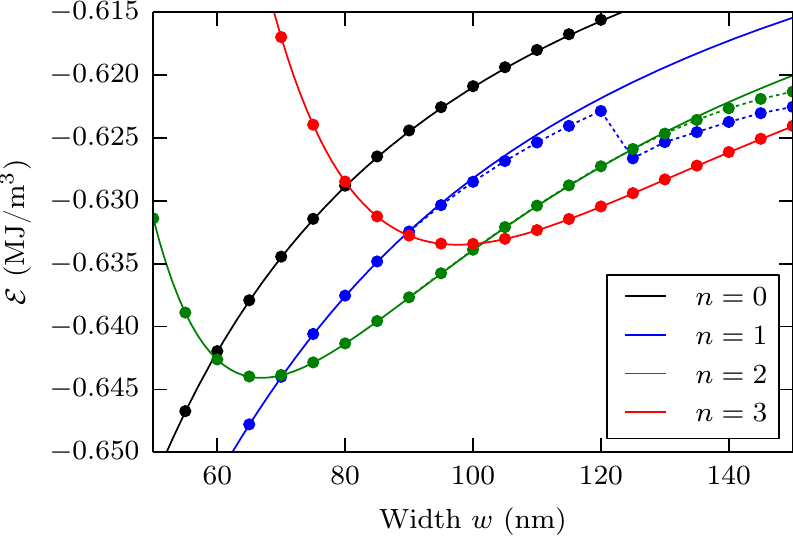}
\caption{\label{fig:cycloiddeform} The energy densities of the numerically relaxed cycloidal states $n=0,1,2,3$ in thin strips with DMI strength~$D=4$mJ/m$^2$ are shown by dots. Lines show the analytical results, previously plotted in Fig.~\ref{fig:cycloidgroundstates}.}
\end{figure}

\begin{figure}[h]
\includegraphics{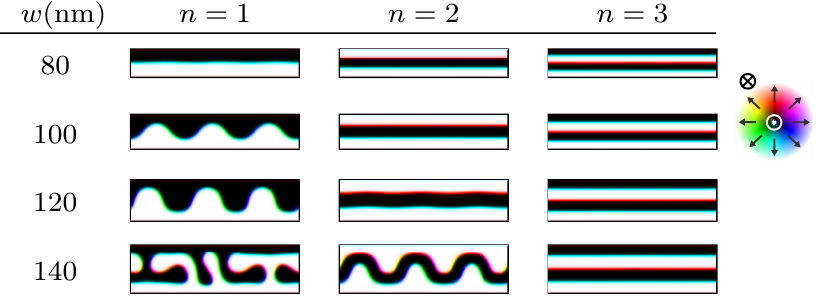}
\caption{\label{fig:selected} The cycloidal states $n=1,2,3$ in thin strips with $D=4$mJ/m$^2$ after relaxation using micromagnetic simulations. The direction of the magnetization is depicted by colors shown in the color wheel. }
\end{figure}

In this section we investigate possible deformations of the cycloidal state in a PMA strip when we drop the assumptions that the magnetic state is constant along the length of the strip and that the magnetization direction lies in the $(x,z)$-plane. Analytic calculations are no longer possible, and we resort to micromagnetic simulations. Periodic boundary conditions and a large simulation box (500~nm) are used in the $y$-direction in order to simulate an infinitely long strip. The initial states in these simulations are cycloidal with a small amount of random noise in order to trigger possible deformations. The resulting energy densities after relaxation of the cycloidal states $n=0,1,2,3$ for DMI strenght~$D=4$mJ/m$^2$ are shown in Fig.~\ref{fig:cycloiddeform}. Examples of the obtained magnetic states are shown in Fig.~\ref{fig:selected}.

The results for the quasi uniform state ($n=0$) correspond exactly with the analytic results. The same is true for the cycloidal state $n=1$ if the width of the strip is small, i.e.~$w<80$~nm. If the width of the strip is larger ($w>80$~nm), we observe buckling in the domain wall, which somewhat lowers the energy density. The magnetization is no longer constant along the $y$-direction. If the width of the strip is taken even larger, $w>125$~nm, the energy density drops drastically after a complex deformation of the initial state. Note that for $w=140$~nm, the typical domain width in the relaxed $n=1$ state is similar to the domain widths in the cycloidal state $n=3$ (ground state). Further increasing the width of the strip will yield similar results for the cycloidal states $n>1$. For example, note the buckled domains in the relaxed cycloidal state $n=2$ in a 140~nm wide strip in Fig.~\ref{fig:selected}.

The relaxation of \emph{perfect} cycloidal states (without buckling) can result in magnetic states that are no longer perfect cycloids. However, all the observed ground states are perfect cycloids, and are thus analytically calculable. The analytic results agree perfectly with the numerical results of the uniform state and of the cycloidal states without buckling (see Fig.~\ref{fig:cycloiddeform}). This justifies the approximation of the demagnetization of thin strips in analytic calculations by using the effective anisotropy~$K_{\mathrm{eff}}$. This is not surprising since the thickness of the strip is much smaller than its width.

\section{\label{sec:square}Square platelets}

In this section we study the ground state and the excited states of square mesoscopic PMA platelets as a function of the side length~$l$ and the DMI strength~$D$. Relaxing a randomly magnetized sample using the LLG equation gives one of the stable states. Repeating this process for different initial magnetic states, sizes~$l$, and DMI strengths~$D$ will reveal the full phase diagram of square PMA platelets. The number of stable states in mesoscopic samples with a low DMI strength or a small side length ($l<60$~nm) will turn out to be limited, which facilitates identifying the ground state as well as all excited states. This is done in Sec.~\ref{sec:smallsquare}. If the platelet is large ($l>60$~nm) and the DMI is strong, then the number of possible states can be very large, making it difficult to identify all stable states. However, it is still possible to determine the ground state. This is detailed in Sec.~\ref{sec:largesquare}.

\subsection{\label{sec:smallsquare}Excited states in small platelets}

\begin{figure}[h!]
\includegraphics{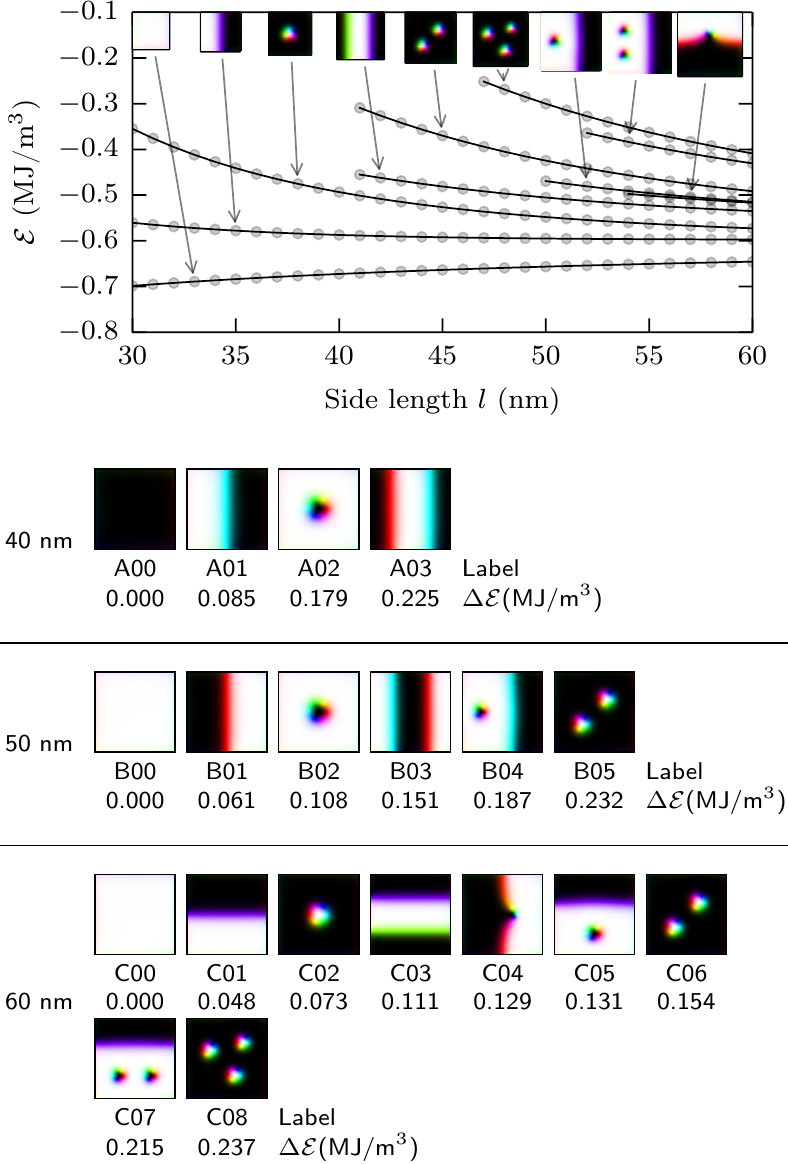}
\caption{\label{fig:d3} The energy densities~$\mathcal{E}$ of the magnetic states of a square $l\times l$ platelet with DMI strength~$D=3$~mJ/m$^2<D_c$. The magnetic states for platelets with side lengths 40~nm, 50~nm and 60~nm are shown separately and labeled in order of their energies. The energy difference with the quasi uniform state, shown below the label, is given in MJ/m$^3$.}
\end{figure}

\begin{figure}[h!]
\includegraphics{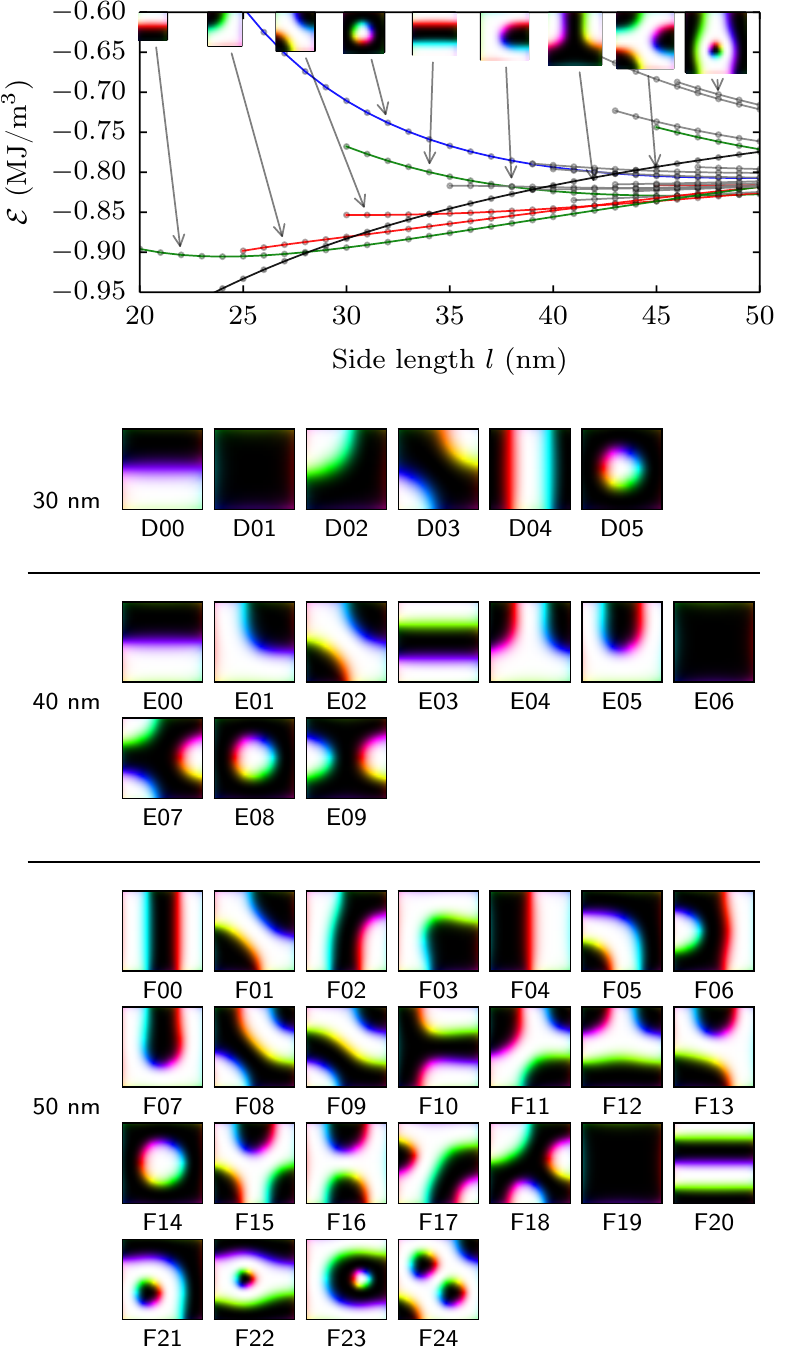}
\caption{\label{fig:d5} The energy densities~$\mathcal{E}$ of the magnetic states of a square $l\times l$ platelet with DMI strength~$D=5$~mJ/m$^2>D_c$: the uniform state (black), the parallel cycloidal states (green), diagonal cycloidal states (red), the single skyrmion state (blue), and other states (gray). The magnetic states for platelets with side lengths 30~nm, 40~nm and 50~nm are shown separately and labelled in the order of their increasing energies.}
\end{figure}

We identified all stable states in square platelets with a size length below 60~nm for DMI strengths $D=3$~mJ/m$^2<D_c$ and $D=5$~mJ/m$^2>D_c$. To convince ourselves that we identified every possible state, we used 10000 initial configurations for each set of parameters, while a few hundred initial configurations are usually sufficient to find all states in such small platelets. This brute force method yields many equivalent states, where we took a single representative state for each set of equivalent states using a comparison algorithm. The states are compared pairwise, taking into account the $D_{4h}$ symmetry of the sample. The representative states and their energies are shown in Fig.~\ref{fig:d3} for $D=3$~mJ/m$^2$ and in Fig.~\ref{fig:d5} for $D=5$~mJ/m$^2$. Some of the representative states are labeled for convenient referral.

Figure~\ref{fig:d3} shows that a stable magnetic state in a platelet of certain size is not necessarily stable in smaller platelets. For example, the excited state C08 shown in Fig.~\ref{fig:d3} is unstable in square platelets with a side length smaller than 47~nm, as one of the three skyrmions will be pushed out of the sample.

For weak DMI strengths $D<D_c$ ($D=3$~mJ/m$^2$ in our case), the ground state in platelets of arbitrary size is uniform. Furthermore, the sequence of the excited states ordered by their energies does not depend on the size of the platelet. The excited states contain distinct features such as skyrmions and domain walls, which can be considered as particle-like excitations. The creation of a domain wall or skyrmion will generally increase the energy. However, this energy difference is not trivial. For instance, the energy difference between the double skyrmion state C06 and the uniform state C00 is not twice as large as the energy difference between the single skyrmion state C02 and the uniform state C00. The same holds true for states with domain walls or with the combination of skyrmions and domain walls. We thus infer that an important ingredient is the repulsion energy between skyrmions, domains, and boundaries. Further, a special kind of domain wall is identified in C04. This domain wall is an ordinary N\'eel wall except at the center, where the in-plane magnetization makes a full rotation. The topological charge of this state is one due to this rotation, just as in a skyrmion.

The phase diagram of the platelets becomes more complex for increasing DMI strengths~$D$. Figure~\ref{fig:d5} shows that the number of possible excited states can be very large for DMI strengths larger than $D_c$. For $D=5$~mJ/m$^2$, we identified 25 and 77 different stable states in square platelets with, respectively, a side length of $l=50$~nm and $l=60$~nm. Identifying all possible stable states for larger films is a very laborious task. Furthermore, the sequence of the magnetic states ordered by their energies does depend on the size of the platelet, which also contributes to the complexity of the phase diagram.

The ground states of square platelets shown in Fig.~\ref{fig:d5} are cycloids parallel with an edge. The number of domain walls in the cycloid depends on the size of the platelet. For example, for $l=40$~nm there is a single domain wall in the ground state E00, while for $l=50$~nm there are two domain walls in the ground state F00. This is consistent with our analytical calculations. In general, the low energy states have a cycloidal character. For instance, state E02 can be considered as a cycloidal state parallel with the diagonal of the square platelet, and state F02 can be considered as a slightly deformed cycloid. However, not every cycloidal state has a relatively low energy. Note that, e.g., cycloidal state F20 is a high-energy state due to its small period. Other high-energy states, such as F21-F24, contain skyrmions. These states are stable since the skyrmions are topologically protected.

The effects of the square shape and the boundaries of the studied platelets are visible in Fig.~\ref{fig:d5}. For example, the only difference between states F00 and F02 is caused by the upper right corner. If the effect of the boundary would be weaker or the shape more round, then state F02 would transform into ground state F00 during relaxation without the need to jump over an energy barrier. Other sets of states in which this effect is visible are \{F11,F12,F13,F15\} and \{F08,F09\}. The energy differences between the states within one set are small compared to the energy differences between states of different sets.

\subsection{\label{sec:largesquare}Ground state of square platelets}

\begin{figure}[h]
\includegraphics{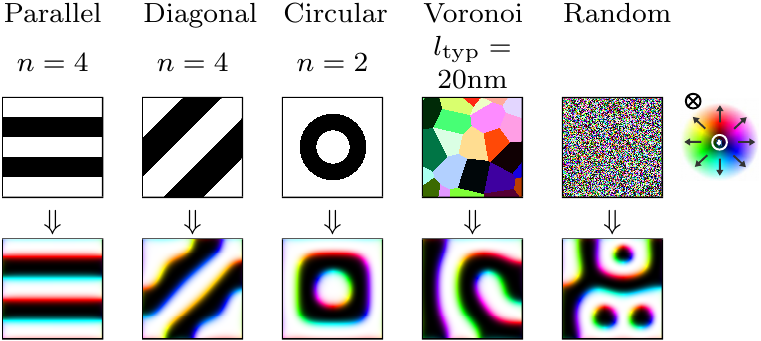}
\caption{\label{fig:initconf} An example of different types of initial magnetization and the resulting relaxed state in square platelets.}
\end{figure}

Using random initial magnetizations in large samples with strong DMI will predominantly yield high-energy states. The reason is that the randomness of the magnetization causes the formation of many small skyrmions, which in turn stabilizes the high energy state since skyrmions are topologically protected. We constructed a coarser random distribution of magnetization by using uniformly magnetized Voronoi domains, in order to avoid the formation of small skyrmions. Varying the size of the Voronoi domains yields a multitude of stable states, with disperse energies. Besides using this coarse random initial magnetization, we also identified some smart choices for the initial state in order to find the low-energy states in large samples. One can imagine that cycloidal states discussed in Sec.~\ref{sec:strip} are good candidates as low-energy states. We thus initialize the calculation from the cycloidal state parallel with an edge of the square platelet (\emph{parallel state}) or parallel with the diagonal (\emph{diagonal state}). We also consider the radially symmetric cycloidal states (\emph{circular state}), which are actually skyrmionic. These different types of initial configurations and some typical results are shown in Fig.~\ref{fig:initconf}.

\begin{figure}[h]
\includegraphics{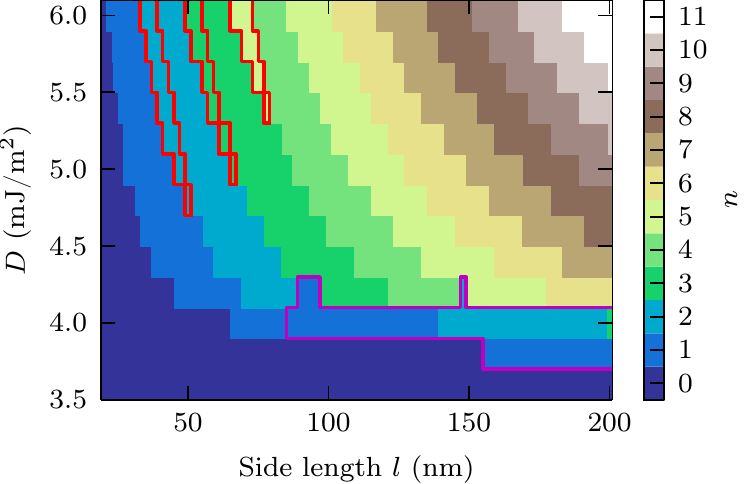}
\caption{\label{fig:groundsquare} The ground states of a square $l\times l$ platelet. Different cycloidal states are represented by $n$ (as defined in section~\ref{sec:analytic}). The states are diagonal inside the red borders, circular inside the magenta border and parallel elsewhere. The stepwise character of the delimiting lines is a side effect of the finite resolution of the phase diagram (2~nm\,$\times$\,0.2~mJ/m$^2$).}
\end{figure}

After selecting the lowest energy state for different side lengths~$l$ and DMI strengths~$D$ of the samples, we obtain the phase diagram shown in Fig.~\ref{fig:groundsquare}. The ground state is always, as already suspected, a parallel, diagonal or a circular state. In most cases, the ground state is a parallel state, which is very similar to the cycloidal states in an infinite strip. Consequently, the phase diagram shares general trends with the phase diagram of the magnetic state in an infinitly long strip shown in Fig.~\ref{fig:cycloidgroundstates}(c). However, there are two important differences. Firstly, the ground state in large samples is circular (skyrmionic) in the vicinity of the critical DMI strength $D_c$. Secondly, there are regions in the phase diagram where the ground state is diagonal. This can be explained by pointing out that, in comparison with the parallel state, the period of a diagonal state in one of these regions is closer to the period of the cycloid in an infinite film.

We end this discussion by mentioning that the skyrmion (or double-wall skyrmion) in the circular ground state around the critical DMI strength~$D_c$ is deformed to a rounded square in large platelets. This gives the state a cycloidal character in both directions of the sample symmetry, and the periods of the cycloids are maximized. For large films, the energy contribution of the relatively small rounded corners becomes negligible. This explains why the circular state is the ground state in large platelets with a DMI strength close to $D_c$.

\section{\label{sec:conclusions}Conclusions}

We have investigated in depth the magnetic phase diagram of thin strips with perpendicular magnetic anisotropy and in the presence of Dzyaloshinskii-Moriya interaction (DMI). We have started the analysis by showing how the cycloidal states in such mesoscopic strips can be calculated analytically, and how confinement promotes hysteretic effects and excited magnetic states. We further resort to micromagnetic simulations, to show that numerical results agree very well with the analytic model for a non-stretched cycloidal state. On the other hand, a stretched cycloidal state is shown to buckle in the numerical experiments, and will deform drastically in order to minimize the energy.

To address further the confinement effects on the magnetic state in chiral mesoscopic magnets, we reported the detailed phase diagram for square platelets. We show that the excited magnetic states in square samples with a weak DMI ($D<D_c=4\sqrt{AK_{\mathrm{eff}}}/\pi$) consist of well-defined skyrmions and domain walls. We find that the energy of a domain wall across the sample is lower than the energy of a skyrmion independently of the sample size, and that stable states with increasing energy sequentially comprise one added wall or skyrmion, all of which interact repulsively to form a stable configuration.

In the case of a strong DMI ($D>D_c$), the phase diagram is very complex. Besides the known skyrmion-skyrmion and skyrmion-edge interactions, we point out the interaction of the domain wall with sample edges (connecting the adjacent or the facing edges of the sample), while interacting with the present skyrmion(s) as well. As a general rule, the cycloidal states with domain walls parallel to the sample edge have the lowest energy, followed by the cycloidal states with diagonal domains, and then those comprising skyrmions. This rule deviates only in large platelets with DMI close to the critical value ($D\approx D_c$), where we have observed skyrmionic ground states.

All together, we emphasize the potential of mesoscopic chiral magnets (with different outer geometry, or with engineered cavities) to stabilize skyrmionic and hybrid skyrmionic-cycloidal states that are otherwise unattainable. Interactions of those states with strategically applied spin-current, and magnetic field, are yet to be explored. Control of transitions between those rich states can enable multibit, nonvolatile magnetic storage, while magnon scattering and intereference between different constituents in those states is worthy of further investigation in this rapidly growing field of mesoscopic physics.


\bibliographystyle{apsrev4-1.bst}
\bibliography{bibtex,coey}

\end{document}